\documentclass[aps,pra,superscriptaddress,twocolumn]{revtex4}
\usepackage{natbib} 
\usepackage{float}
\usepackage{bm}
\usepackage{graphicx,amsmath,latexsym,amssymb}
\usepackage{color}
\usepackage{dcolumn}

\begin{document}
\title{
Nonreciprocal and nonconservative forces on binary systems of identical atoms}
\author{J. S\'anchez-C\'anovas} 
\affiliation{Departamento de F\'isica Te\'orica, At\'omica y \'Optica,  Universidad de Valladolid, Paseo Bel\'en 7, 47011 Valladolid, Spain}
\author{M. Donaire} 
\email{manuel.donaire@uva.es}
\affiliation{Departamento de F\'isica Te\'orica, At\'omica y \'Optica,  Universidad de Valladolid, Paseo Bel\'en 7, 47011 Valladolid, Spain}

\begin{abstract}
The dynamical and radiative features of an excited system of two identical atoms are analysed. The metastability of the system, the directionality of its emission  and its internal forces are studied. Closed-form expressions are derived for the time-evolution of the system, for the angular distribution of its spontaneous emission, and for its internal dipole forces, both conservative and nonconservative. The latter reveals the presence of nonreciprocal forces, which leads to a net oscillatory force upon the system. We estimate that, for a free binary system of Li Rydberg atoms, the net internal force may cause a displacement of its center of mass as large as $120 nm$ over a lifetime.
\end{abstract}
\maketitle

\section{Introduction}  
 
The collective behaviour of an excited system of  identical two-level molecules was first studied by Dicke \cite{Dicke} and has since become a subject of interest in different contexts of optics and quantum computing \cite{Cirac1, Cirac2, Haroche1, Haroche2,Reinhard2007}. The most relevant feature of such a system is its coherent evolution in time, together with its exceptional radiative properties due to collective effects. Thus, inhibition and enhancement of spontaneous emission is achieved for highly symmetric configurations of the atomic states in dense ensembles.  The so-called Dicke model of two identical two-level atoms has been profusely studied in the literature \cite{Stephen,Craigbook,Varfolomeev1,Varfolomeev2}. The model has been later extended to a number of analogous physical systems \cite{artificial_atoms,dots}, where the attention has focused on their radiative properties in the search for  superradiant and subradiant states. 

Generally speaking, the Dicke model for atoms belongs to the broader area of research of collective effects and dispersion forces in binary systems of neutral atoms  \cite{vdW,WileySipe,Milonnibook, BuhmannbookI,BuhmannbookII,Craigbook,BuhmannScheel}. Both phenomena result from the atomic correlations induced by
the coupling of the quantum fluctuations of the electromagnetic (EM) field in its vacuum state with the dipole fluctuations of the atomic charges in stable or metastable states. For the case of atoms in their ground states the atomic interactions are generally referred to as van der Waals interactions, and the forces which derive from them are computed applying  the usual techniques of stationary quantum perturbation theory \cite{Miltonbook,BuhmannbookI,BuhmannbookII,Craigbook,vdW,Polder}. 
\begin{figure}
	\begin{center}
  		\includegraphics[width=85mm,angle=0,clip]{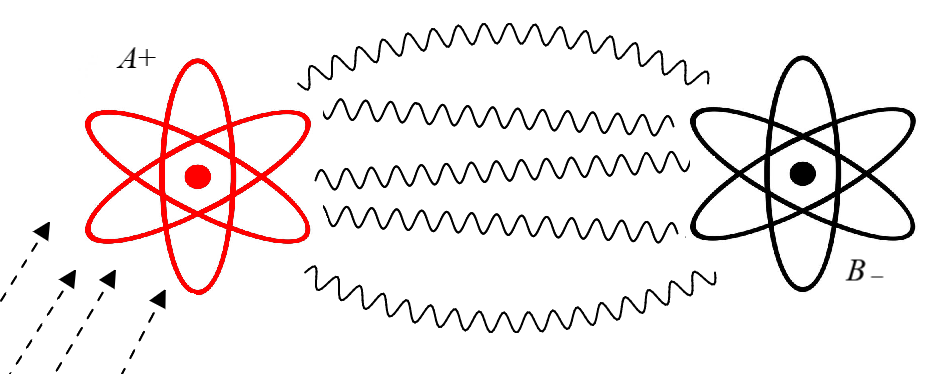}
 		\caption{Pictorial illustration of the initial setup of the system.}
	\end{center}
\end{figure}
In contrast, for the case of excited atoms, their interactions contain resonant contributions. The latter are associated to resonances in the integrals over the normal modes of the virtual photons which mediate the interactions. It has been recognised in the last decade that a time-dependent approach is necessary to account appropriately for the incoherent dynamics inherent to metastable states  \cite{Passante,Berman,MePRL,Milonni,Pablo,MePRAvdW,Sherkunov1,Sherkunov2}. Nonetheless, for the case of the interaction between two dissimilar excited atoms, the adiabatic approximation is generally sufficient to compute forces and energies. The reason being that the excitation of the atoms is adiabatic with respect to the internal dynamics of the atomic system, whose typical time scale is determined by the detuning between the transition frequencies of both atoms. Hence, the interaction between dissimilar atoms is proportional to the fourth power of the transition dipole moments and inversely proportional to the detuning.

However, for the case of a binary system of excited and identical atoms, the problem becomes inherently time-dependent Ref.\cite{Julio_PRA1}. This so because, on the one hand, the system is degenerate and an adiabatic excitation is not feasible. In fact, a sudden excitation is generally a good approximation 
to model the preparation of the initial state of the system. On the other hand, that excited state is highly non-stationary since its dynamics comprises both the coherent transfer of the excitation between the atoms and its incoherent decay through spontaneous emission.

In two previous publications \cite{Julio_PRA1,Julio_PRA2} we have shown that, in the weak-interaction regime, the electric dipole interaction between two identical two-level atoms, with one of them initially excited, gives rise to non-reciprocal and non-conservative forces. The apparent violation of the Newtonian principle of action-reaction and total momentum conservation is explained in terms of the linear momentum carried off by the virtual photons which mediate the interaction. At leading order, the resultant forces  are proportional to the square of the transition dipole moments and linear in time, thus much stronger than their analogue in the case of dissimilar atoms. Using realistic values for the atomic parameters, it was estimated that the strength of the forces is too small to be detected experimentally. 

It is the aim of the present article to extend the approach of Refs.\cite{Julio_PRA1,Julio_PRA2} to the calculation of the internal forces and the spectrum of emission in a binary system of identical two-level atoms, with one them initially excited, including non-perturbative effects. Thus, we compute the time-evolution the population levels; the spectrum of emission together with the directionality of this emission; and the non-reciprocal and non-conservative forces upon each atom.
The estimated values of all these effects are enlarged by the degenerate nature of the system, facilitating this way their experimental detection. Hence, we will see that the forces are proportional to the square of the transition dipole moments and oscillate in time, which makes them much stronger than the forces between dissimilar atoms. Finally, this work is intended to pave the way for the estimate of these quantities in a binary systems of Rydberg atoms. The extraordinary properties of these atoms in regards to the strength of their electric dipole moment transition elements and to their lifetimes are expected to yield results which are accessible with current experimental techniques. 

The paper is organized as follows. In Sec.\ref{II} we present the fundamentals of the time-dependent approach used in this paper. In Sec.\ref{III} we derive the time evolution operator of the system, incorporating all the interaction processes within it. In Sec.\ref{IV} we calculate the conservative and non-conservative internal forces, distinguishing between reciprocal and non-reciprocal components. The directionality of the spontaneous emission is analysed. Finally, the conclusions are summarized in Sec.\ref{V}, along with an estimate of the net displacement caused by the non-reciprocal forces on a binary system of Rydberg atoms.
	
\section{Fundamentals}\label{II}
Let us consider a binary system of identical two-level atoms, $A$ and $B$, located a distance $\mathbf{R}$ apart. Each atom, when isolated in free space, possesses a resonant frequency  $\omega_0$ and a linewidth $\Gamma_0$ associated to the transition between the excited and ground states, $|(A,B)_{-}\rangle$ and $|(A,B)_{+}\rangle$, respectively. At time $t=0$ one of the atoms, say atom $A$, is excited by a $\pi$-pulse of a laser  which is tuned with the transition frequency.
Thus,  at time $t=0$, the atomic system is in its ground state, while the electromagnetic (EM) field state is that corresponding to an external laser field of intensity $\epsilon_{0}cE_{L}^{2}$, with $E_{L}$ being the amplitude of the laser electric field. Therefore, the  initial state of the system atom-EM field reads $|\Psi(0)\rangle=|A_-\rangle \otimes |B_-\rangle \otimes |N_{\gamma}\rangle$, where $|N_{\gamma}\rangle$ is the EM state made of a density $N_{\gamma}/\mathcal{V}=\epsilon_{0}E_{L}^{2}/\hbar\omega_{0}$ of laser photons.

 We assume strong coupling between the  laser and the binary system such that the atom gets excited after a time interval $\pi/\Omega_{L}$. In this expression, $\Omega_{L}$ is the Rabi frequency associated to the laser field, $\Omega_{L}=\langle A_+|\mathbf{d}_A|A_-\rangle\cdot\mathbf{E}_{L}$, with $\mathbf{d}_A$ being the transition dipole moment operator of atom $A$. Strong coupling with the laser implies that the atom-laser interaction is much stronger than that between the atoms and $\Omega_{L}\gg\Gamma_{0}$.

Within the framework of Sch\"odinger's picture, at any given time $T>0$ the state of the two-atom-EM field system can be written as $|\Psi(T)\rangle= \mathbb{U}(T)|\Psi(0)\rangle$, where $\mathbb{U}(T)$ denotes the time propagator,
\begin{equation}
	\mathbb{U}(T)= \text{T}-\exp \Big\{-\frac{i}{\hbar} \int_{0}^{T} dt H \Big\}.
\end{equation}

In this equation $H$ is the total Hamiltonian of the system, which can be decomposed as
\begin{equation}
	H=\mathcal{T}+H_A+H_B+H_{EM}+W,
\end{equation}
where $\mathcal{T}=m|\dot{\mathbf{R}}_{A}|^{2}/2+m|\dot{\mathbf{R}}_{B}|^{2}/2$ is the kinetic energy of the centers of mass of the atoms, 
with $m_{A,B}$ being their the masses and $\mathbf{R}_{A,B}$ their position vectors. $H_{A,B}$ are the Hamiltonians of the internal degrees of freedom of each atom, 
$H_{A,B}=\hbar\omega_{0}|(A,B)_{+}\rangle\langle (A,B)_{+}|$; and the Hamiltonian of the free EM field is $H_{EM}=\sum_{\mathbf{k},\boldsymbol{\epsilon}}\hbar\omega(a^{\dagger}_{\mathbf{k},\boldsymbol{\epsilon}}a_{\mathbf{k},\boldsymbol{\epsilon}}+1/2)$,
where $\omega=ck$ is the photon frequency, and the operators $a^{\dagger}_{\mathbf{k},\boldsymbol{\epsilon}}$ and $a_{\mathbf{k},\boldsymbol{\epsilon}}$ are the creation and annihilation operators of photons with momentum $\hbar\mathbf{k}$ and polarization. 


Finally, the interaction 
Hamiltonian in the electric dipole approximation reads $W=W_{A}+W_{B}$, with 
\begin{eqnarray}
	W_{A,B}&=&-\mathbf{d}_{A,B}\cdot\mathbf{E}(\mathbf{R}_{A,B}),\label{WAB}
\end{eqnarray}
where $\mathbf{E}(\mathbf{R}_{A,B})$ are the quantum 
electric field operators at the location of each atom in Schr\"odinger's representation. For the sake of completeness we write the electric and magnetic field operators in terms of the EM vector potential in the Coulomb gauge $\boldsymbol{\nabla} \cdot \mathbf{A}(\mathbf{r},t)=0$, 
\begin{equation}
	\mathbf{A}(\mathbf{r},t)=\sum_{\mathbf{k},\boldsymbol{\epsilon}}\sqrt{\frac{\hbar}{2\omega\mathcal{V}\epsilon_{0}}}
	[\boldsymbol{\epsilon}a_{\mathbf{k},\boldsymbol{\epsilon}}e^{i(\mathbf{k}\cdot\mathbf{r}-\omega t)}
	+\boldsymbol{\epsilon}^{*}a^{\dagger}_{\mathbf{k},\boldsymbol{\epsilon}}e^{-i(\mathbf{k}\cdot\mathbf{r}-\omega t)}].\nonumber
\end{equation} 
That is, the electric and magnetic fields, $\mathbf{E}(\mathbf{R}_{A,B})=-\partial_{t}\mathbf{A}(\mathbf{R}_{A,B},t)|_{t=0}$, 
$\mathbf{B}(\mathbf{R}_{A,B})=\boldsymbol{\nabla}_{A,B}\times\mathbf{A}(\mathbf{R}_{A,B})|_{t=0}$, can be written as sums over normal modes as
\cite{Milonnibook,Craigbook}
\begin{align}\label{AQ}
	\mathbf{E}&(\mathbf{R}_{A,B})=\sum_{\mathbf{k}} \left[\mathbf{E}^{(-)}_{\mathbf{k}}(\mathbf{R}_{A,B})+\mathbf{E}^{(+)}_{\mathbf{k}}(\mathbf{R}_{A,B})\right]\nonumber\\
	&=i\sum_{\mathbf{k},\boldsymbol{\epsilon}}\sqrt{\frac{\hbar ck}{2\mathcal{V}\epsilon_{0}}}
	[\boldsymbol{\epsilon}a_{\mathbf{k},\boldsymbol{\epsilon}}e^{i\mathbf{k}\cdot\mathbf{R}_{A,B}}-\boldsymbol{\epsilon}^{*}a^{\dagger}_{\mathbf{k},\boldsymbol{\epsilon}}e^{-i\mathbf{k}\cdot\mathbf{R}_{A,B}}],\nonumber\\
	\mathbf{B}&(\mathbf{R}_{A,B})= \sum_{\mathbf{k}} \left[\mathbf{B}^{(-)}_{\mathbf{k}}(\mathbf{R}_{A,B})+\mathbf{B}^{(+)}_{\mathbf{k}}(\mathbf{R}_{A,B})\right]\nonumber\\
	&=i\sum_{\mathbf{k},\boldsymbol{\epsilon}}\sqrt{\frac{\hbar}{2ck\mathcal{V}\epsilon_{0}}}\mathbf{k}\times
	[\boldsymbol{\epsilon}a_{\mathbf{k},\boldsymbol{\epsilon}}e^{i\mathbf{k}\cdot\mathbf{R}_{A,B}}-\boldsymbol{\epsilon}^{*}a^{\dagger}_{\mathbf{k},\boldsymbol{\epsilon}}
	e^{-i\mathbf{k}\cdot\mathbf{R}_{A,B}}],\nonumber
\end{align}
where $\mathcal{V}$ is a generic volume and $\mathbf{E}^{(\mp)}_{\mathbf{k}}$, $\mathbf{B}^{(\mp)}_{\mathbf{k}}$ denote the annihilation/creation electric and 
magnetic field operators of photons of momentum $\hbar\mathbf{k}$, respectively.  
For further purposes, we will consider the quadric fluctuations of the electric and magnetic fields in vacuum, which can be expressed in terms of Green's tensors. $\mathbb{G}(kR)$ is the Green tensor of the electric field induced at a distance $\mathbf{R}$ by a point-like dipole of frequency $\omega=ck$,
\begin{equation}
	\mathbb{G}(kR)= -\frac{k e^{i kR}}{4\pi} \left[ \frac{\alpha}{k R}+ \frac{i\beta}{(kR)^2}- \frac{\beta}{(kR)^3} \right],
\end{equation}
where $\alpha= \mathbb{I}-\hat{\mathbf{R}}\hat{\mathbf{R}}$ and $\beta=\mathbb{I}-3\hat{\mathbf{R}}\hat{\mathbf{R}}$ with $\hat{\mathbf{R}}=\frac{\mathbf{R}}{R}$. In terms of $\mathbb{G}(kR)$, the electric field fluctuations in vacuum read
\begin{equation}
	\int d\Theta_{\mathbf{k}}\langle0_{\gamma}|\mathbf{E}^{(-)}_{\mathbf{k}}(\mathbf{R})\:\mathbf{E}^{(+)}_{\mathbf{k}}(\mathbf{0})|0_{\gamma}\rangle=-\frac{8 \pi^2 \hbar c}{\epsilon_0} \,\ \textrm{Im}\:\mathbb{G}(kR),\label{GpaE}
\end{equation}	
 where $|0_{\gamma} \rangle$ is the EM vacuum state. 
On the other hand, $\boldsymbol{\nabla}\wedge\mathbb{G}(kR)$ is the magnetic field created by a point-like electric dipole at a distance $\mathbf{R}$ and frequency $\omega=ck$. In terms of it, 
\begin{equation}
	\int d\Theta_{\mathbf{k}}\langle0_{\gamma}|\mathbf{B}^{(-)}_{\mathbf{k}}(\mathbf{R})\:\mathbf{E}^{(+)}_{\mathbf{k}}(\mathbf{0})|0_{\gamma}\rangle=-\frac{8 \pi^2 i\hbar }{\epsilon_0 k} \,\ \boldsymbol{\nabla}_{\mathbf{R}}\times\textrm{Im}\:\mathbb{G}\nonumber(kR).\label{GpaB}
\end{equation}

\section{Time evolution of the system}\label{III}

The free and unperturbed  Hamiltonian of the system of which the atomic and EM states are eigenstates is identified with $H_{0}=+H_A+H_B+H_{EM}$, from which we define the free time-propagator $\mathbb{U}_{0}(t)=e^{-iH_{0}t/\hbar}$. Upon its free evolution, under the conditions outlined in the previous section, the time evolution of the system with initial state $|\Psi(0)\rangle=|A_-\rangle \otimes |B_-\rangle \otimes |N_{\gamma}\rangle$ is led by four kinds of distinguishable processes, namely, i) those which involve the emission and absorption of virtual photons from excited states; ii) those which involve the absorption and emission of laser photons; iii) those which involve the exchange of single virtual photons between the atoms, once at a time; iv) those which involve the exchange of pairs of virtual photons between the atoms.  In what follows we will split the computation of the elements of the time evolution propagator $\mathbb{U}$ in several steps. First we  define the free time-propagator; next we will introduce the EM self-interaction of each atom and the action of the laser upon atom $A$; later we will account for the atom-atom interaction; and finally, we will consider the elements with photon states.

\subsection{Hierarchy of the interactions. Laser action}

\begin{figure}
\begin{center}
\includegraphics[width=85mm,angle=0,clip]{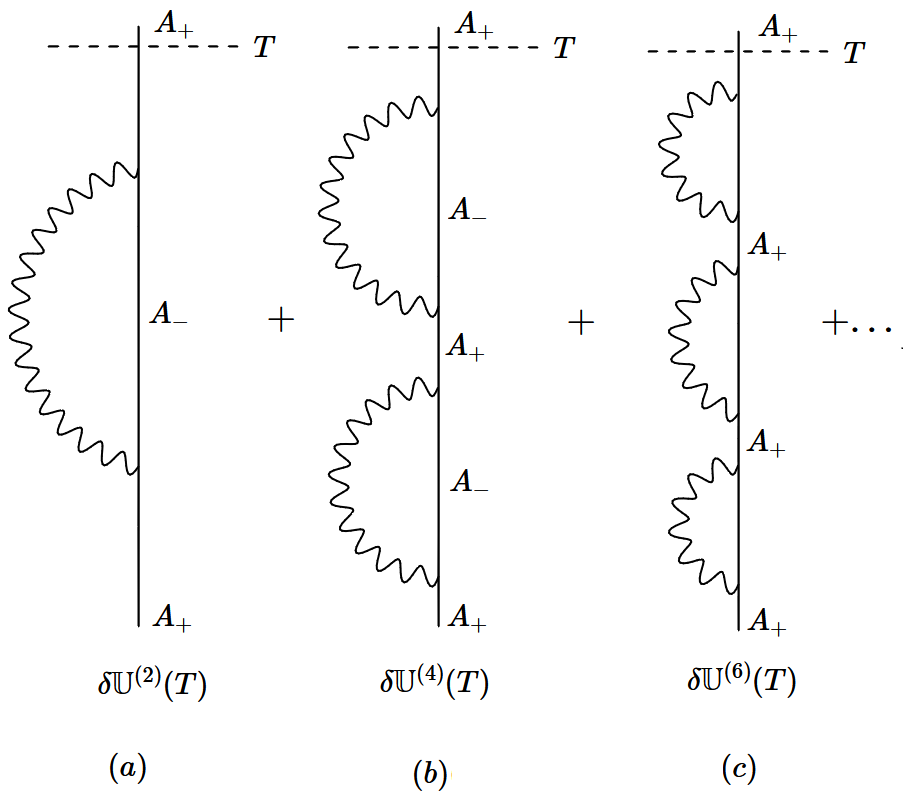}
\caption{Diagrammatic representation of the self-interaction processes contributing to the incoherent attenuation of the $\mathbb{U}_{0}(t)$ element $|A_+\rangle\langle A_+|$ with a factor $e^{-\Gamma_{0}t/2}$. Analogous diagrams hold for the element $|B_+\rangle\langle B_+|$. Note also that $\omega_{0}$ appears here implicitly 'dressed' by the Lamb-shift generated by the same sequence of self-interaction  processes.}\label{fig1}
\end{center}
\end{figure}

\begin{figure}
\begin{center}
\includegraphics[width=85mm,angle=0,clip]{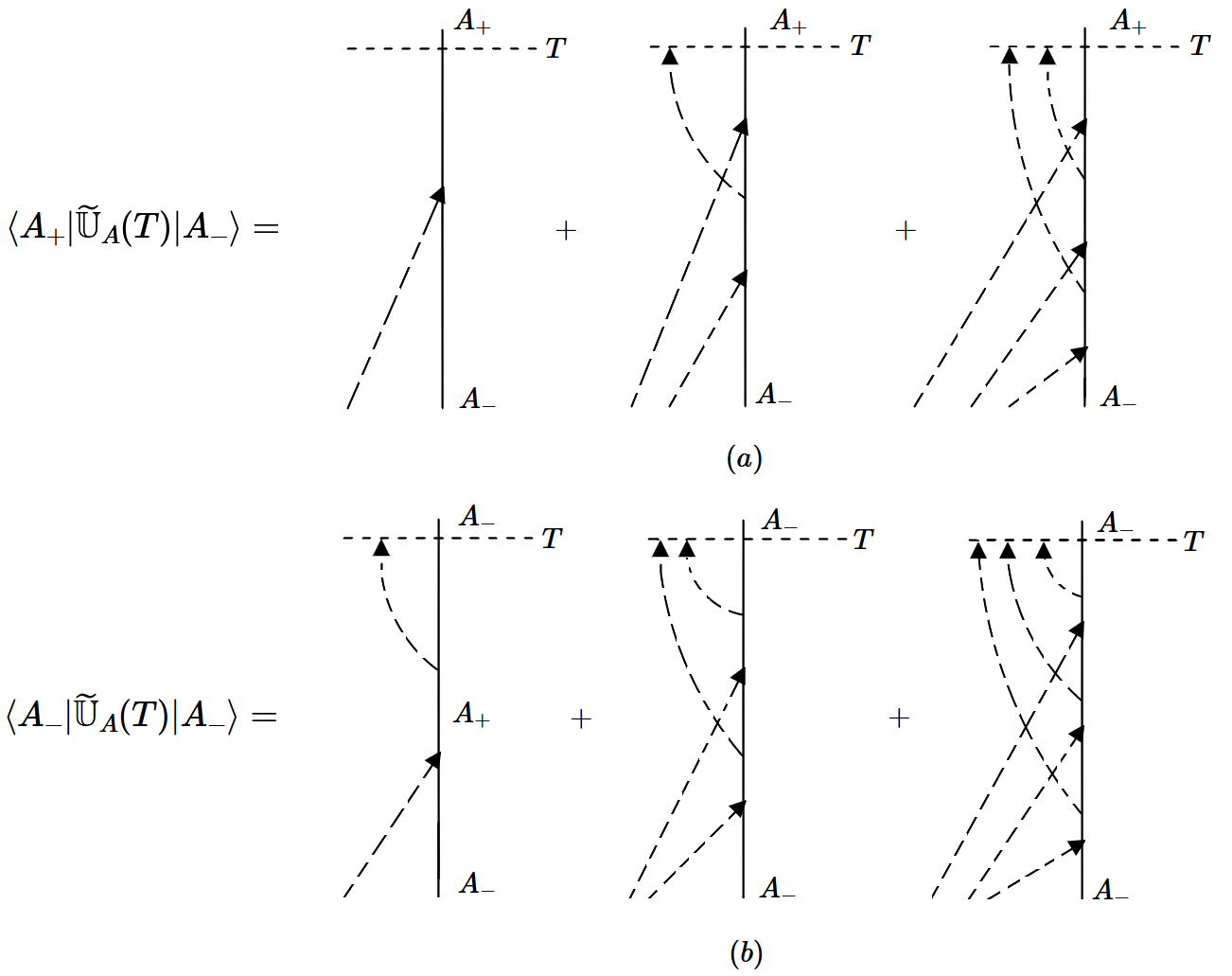}
\caption{Diagrammatic representation of the resonant processes (i.e, RWA) of the laser action upon atom $A$. The contribute to the  $\widetilde{\mathbb{U}}_{A}$ elements $|A_+\rangle\langle A_-|$ [(a)] and $|A_-\rangle\langle A_-|$ elements [(b)], respectively. The laser photons appear depicted by dashed lines. We omit non-interacting laser photons.}\label{fig2}
\end{center}
\end{figure}

\begin{figure}
\begin{center}
\includegraphics[width=85mm,angle=0,clip]{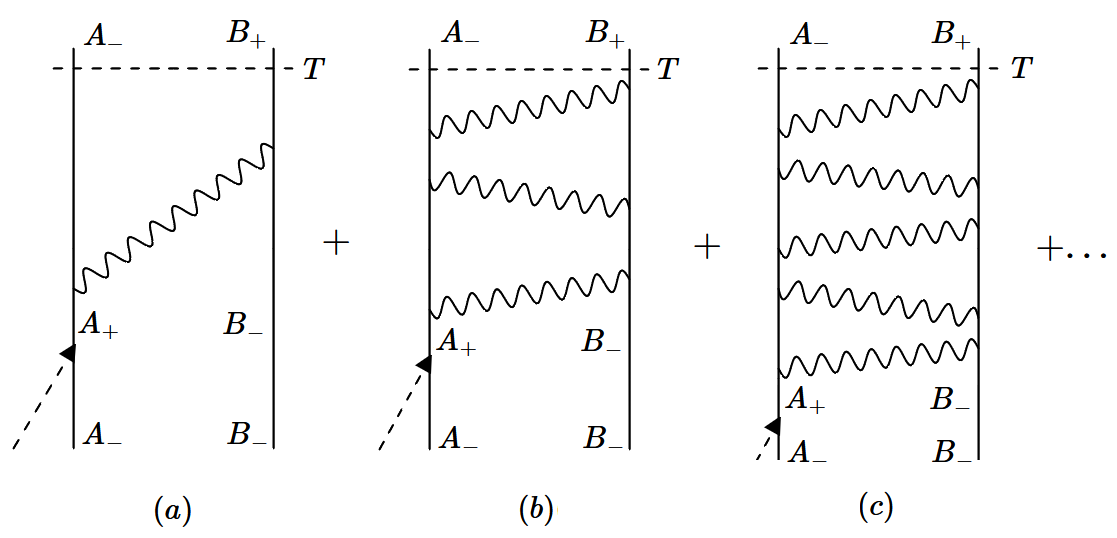}
\caption{Diagrammatic representation of the resonant processes of the interatomic interaction which contribute to $\mathbb{U}_{B_{+}}$, in which an odd number of single photons is exchanged between both atoms. The action of the laser at the bottom of each diagram is equivalent to the sudden excitation approximation in the strong laser coupling regime.}\label{fig3}
\end{center}
\end{figure}

\begin{figure}
\begin{center}
\includegraphics[width=85mm,angle=0,clip]{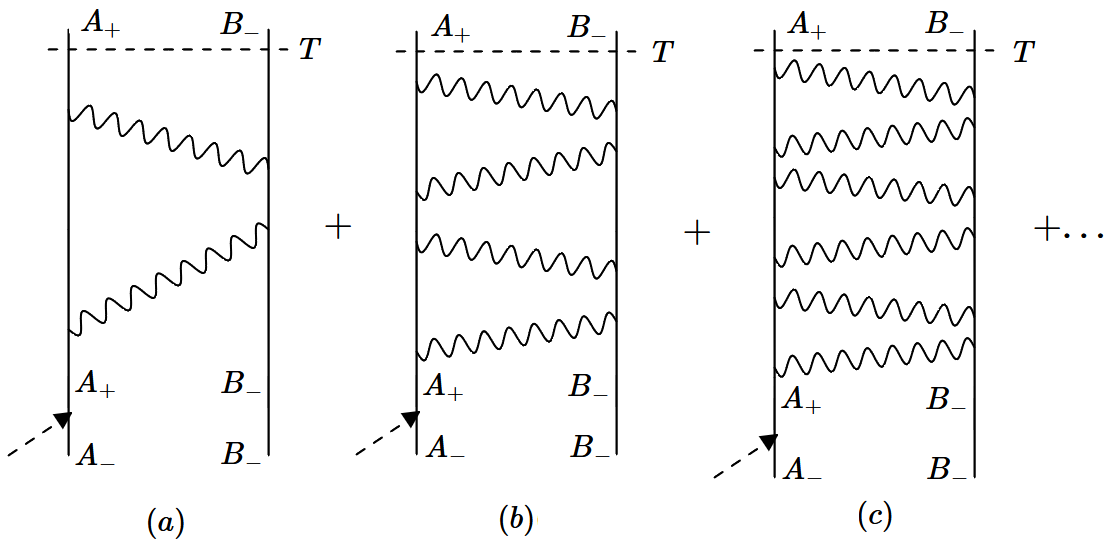}
\caption{Diagrammatic representation of the resonant processes of the interatomic interaction which contribute to $\mathbb{U}_{A_{+}}$, in which an even number of single photons is exchanged between both atoms. The action of the laser at the bottom of each diagram is equivalent to the sudden excitation approximation in the strong laser coupling regime. The dashed arrows at the bottom of each diagram stands for the quasi-sadden excitation of atom $A$ by the action of the external laser.}\label{fig4}
\end{center}
\end{figure}

\begin{figure}
\begin{center}
\includegraphics[width=85mm,angle=0,clip]{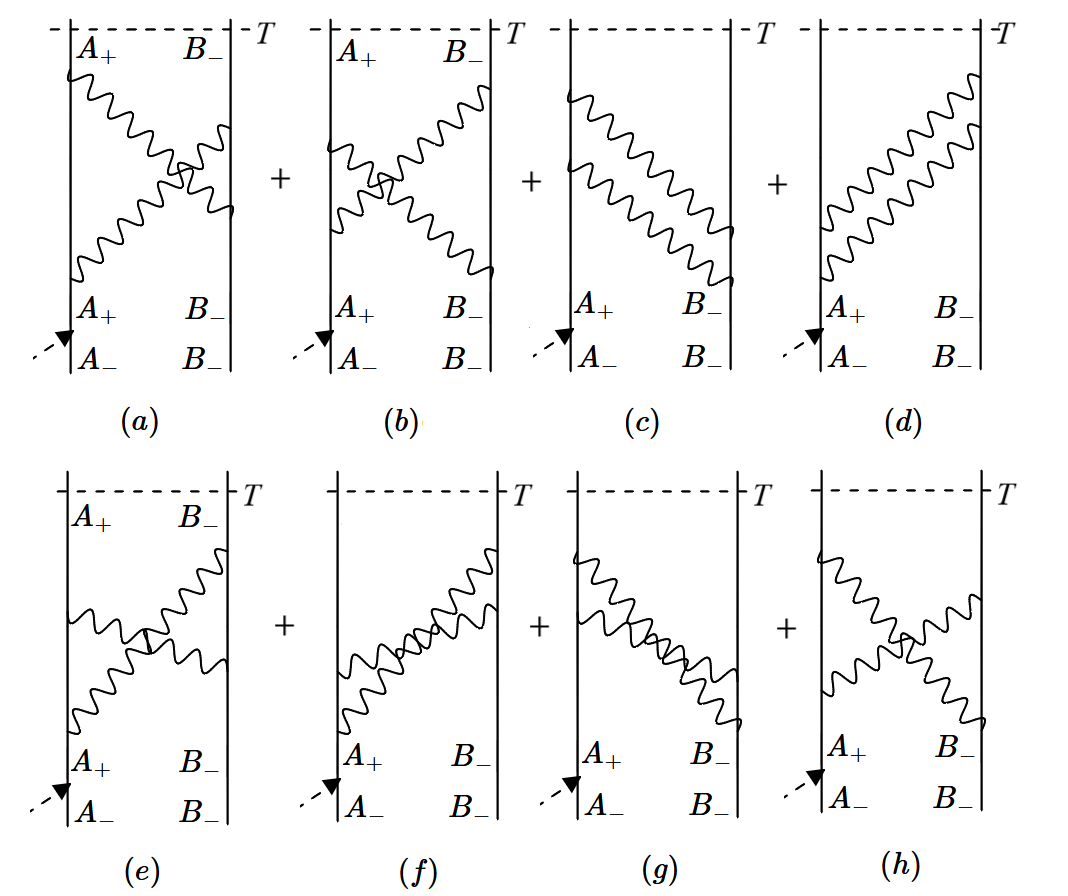}
\caption{Diagrammatic representation of the off-resonant processes of the interatomic interaction which contribute to an effective phase-shift of the $\mathbb{U}$ element $|A_+,B_{-}\rangle\langle A_-,B_{-}|$, with two-photon intermediate states. The action of the laser at the bottom of each diagram is equivalent to the sudden excitation approximation in the strong laser coupling regime. The dashed arrows at the bottom of each diagram stands for the quasi-sadden excitation of atom $A$ by the action of the external laser.}\label{fig5}
\end{center}
\end{figure}

Concerning the processes i), they account for the EM self-interaction of the atom, and give rise to the well-known Lamb shift and natural emission, which are depicted in Fig.\ref{fig1}. Amongst the processes ii) , the dominant ones are those which fit into the \emph{rotatory wave approximation} (RWA), i.e., the resonant ones of the sort of those in Fig.\ref{fig2} in which absorption of single photons from the state $|A_-\rangle$ and stimulated emission of single photons from the state $|A_+\rangle$ alternate consecutively. In the processes iii), we can differentiate resonant and off-resonant processes. In the former case, illustrated in Figs.\ref{fig3} and \ref{fig4},  single photons are emitted from one of the atoms in its excited state and are absorbed by the other atom in its ground state, alternatively. They amount to the resonant energy transfer between the atoms. In the latter case, a single photon is emitted from one of the atoms in its excited state and is absorbed by the other atom in its ground state. Finally, the processes iv) are depicted in Fig.\ref{fig5}. They amount to those van der Waals interactions in which intermediate states contain pairs of photons. They include mostly off-resonant components. 

The resonant processes of self-interaction, laser excitation, and resonant energy transfer of the kinds i), ii) and iii), respectively, give rise to summable series, separately. The first ones leads to an exponential attenuation of the wave function of the excited state with an effective prefactor $e^{-\Gamma_{0}t/2}$, with $\Gamma_{0}$ being given by 
\begin{align}
	\Gamma_{0}&=\partial_{T}\int\frac{d^{3}k}{(2\pi)^{3}}\mathcal{V}|\langle A_-,B_-;\gamma_{\mathbf{k},\boldsymbol{\epsilon}}|\mathbb{U}_{0}(T)|A_+,B_-;0_{\gamma}\rangle|^2\nonumber\\
	&=\frac{-2k_{0}^{2}}{\epsilon_{0}\hbar}\boldsymbol{\mu}_{A}\cdot\textrm{Im}\mathbb{G}(k_{0}r)\cdot\boldsymbol{\mu}_{A},\quad
	k_{0}r\rightarrow0^{+},\:\:\Gamma_{0}T\ll1.\label{G0}
\end{align}
In an effective manner, the iterative series of absorption and re-emission processes depicted in Fig.\ref{fig1} is integrated in the evolution propagator attaching an exponential attenuating factor $e^{-\Gamma_{0}t/2}$ to the free  propagator $\mathbb{U}_{0}(t)$ when applied upon the excited atomic states $|A_{+}\rangle,|B_{+}\rangle$. Hereafter, this procedure is implicitly assumed in all the calculations.

The laser excitation involves processes in which the atom $A$ undergoes a series of photon absorptions and emissions induced by the laser field. Amongst them, those which involve alternate events of absorption and stimulated emission of laser photons  constitute the dominant resonant processes --see Fig.\ref{fig2}-- which lead to summable series of the form
\begin{equation}
	\begin{split}
		\widetilde{\mathbb{U}}_{A}(t)&=
		\cos \left( \frac{\Omega_L }{2}t  \right) |A_- \rangle \langle A_- |-i\sin \left( \frac{\Omega_L }{2}t \right) |A_-\rangle \langle A_+|\nonumber\\
		&+e^{- i \omega_0 t-\Gamma_{0}t/2} \cos \left( \frac{\Omega_L }{2}t \right) |A_+ \rangle \langle A_+|\nonumber\\
		&-ie^{- i \omega_0 t-\Gamma_{0}t/2}  \sin \left( \frac{\Omega_L }{2}t \right) |A_+\rangle \langle A_-|,
	\end{split}
\end{equation}
where we omit the laser photons in the state. The rest of the processes, not fully resonant, contribute with terms of order $\Omega_{L}/\omega_{0}$ smaller. Their neglect is equivalent to the RWA. 


\subsection{Interatomic interaction. Atomic dynamics}

On the other hand, the resonant processes of the kind ii) correspond to the interatomic interaction and involve one-photon and no-photon intermediate states.  Because the states $|A_+\rangle \otimes |B_-\rangle \otimes|0_{\gamma}\rangle $, $|A_-\rangle \otimes |B_+\rangle \otimes |0_{\gamma}\rangle$ and $|A_-\rangle \otimes |B_-\rangle \otimes |\gamma_{\mathbf{k},\boldsymbol{\epsilon}}\rangle $ constitute a degenerate subspace, the interaction can only be treated within the framework of time-dependent perturbation theory. Its strength, $\widetilde{\Omega}(R)$,  can be read from the evaluation of the off-diagonal element of the EM interaction Hamiltonian --see Fig.\ref{fig3}(a)-- assuming that the atom $A$ is initially excited. The frequency $\widetilde{\Omega}(k_{0}R)$ is the time derivative of the time evolution operator element from $|A_+B_-\rangle$ to $|A_-B_+\rangle$ at leading order, i.e., at order two in $W$. That is nothing but the corresponding interaction energy element, i.e.,
\begin{align}
\widetilde{\Omega}(k_{0}R)&=\partial_{T}\Bigl[-\hbar^{-2}\langle A_-,B_+|\mathbb{U}_{0}^{\dagger}(T)e^{+\Gamma_{0}T/2}\int_{0}^{T}dt\nonumber\\
&\times\mathbb{U}_{0}(T-t)W\int_{0}^{t}dt' \mathbb{U}_0(t-t')W\mathbb{U}_0(t') |A_+,B_-\rangle\Bigr]\nonumber\\
&=-i\hbar^{-2} \langle A_-,B_+|\mathbb{U}_{0}^{\dagger}(T) W\int_{0}^{T}dt\nonumber\\
&\times \mathbb{U}_0(T-t)W\mathbb{U}_0(t) |A_+,B_-\rangle\nonumber\\
&=\hbar^{-1}\epsilon_{0}^{-1}k_0^2 \boldsymbol{\mu}_{A}\cdot\mathbb{G}(k_0 R)\cdot\boldsymbol{\mu}_{B},
\end{align}
where $\boldsymbol{\mu}_{A}=\langle A_-|\mathbf{d}_{A}|A_{+}\rangle$,  $\boldsymbol{\mu}_{B}=\langle B_-|\mathbf{d}_{B}|B_{+}\rangle$. Hereafter we will use the following notation
\begin{align}
	\widetilde{\Omega}(k_{0}R)&=\hbar^{-1}\epsilon_{0}^{-1}k_0^2 \boldsymbol{\mu}_{A}\cdot \left[\textrm{Re}\mathbb{G}(k_{0} R) + i \textrm{Im}\mathbb{G}(k_{0} R) \right]\cdot\boldsymbol{\mu}_{B}\nonumber\\
	&\equiv \Omega_{k_{0}R}-i\Gamma_{k_{0}R}.
\end{align}

The mixture of the processes contributing to the laser excitation and resonant energy transfer leads, generically, to non-resummable series for the time-evolution propagator. However, in the situation of our interest, the strong laser coupling regime, defined by the condition $\Omega_{L}\gg\Omega_{k_0 R}$, all the processes of the laser excitations can be summed up in the first instance such that, upon considering the interatomic interactions, the unperturbed 'dressed' time evolution operator of the system is $\widetilde{\mathbb{U}}_0(T)=\widetilde{\mathbb{U}}_A(T) \otimes \mathbb{U}_0^{B}(T) \otimes \mathbb{U}_{0}^{EM}(T)$, and  $W$ will be considered hereafter the only perturbation of the system. In this expression, $ \mathbb{U}_0^{B} \otimes \mathbb{U}_{0}^{EM}$, stand for the free time-propagators of the EM field and the atom $B$. Hence, the complete time-propagator can be written as 
\begin{equation}
	\mathbb{U}(T)= \widetilde{\mathbb{U}}_0(T) \,\ \text{T}-\exp{\left[\frac{-i}{\hbar}\int_0^T \widetilde{\mathbb{U}}_0^{\dagger}(T)W(T)\widetilde{\mathbb{U}}_0(T)\right]},
\end{equation}
where $\mathbb{U}(T)=\widetilde{\mathbb{U}}_0(T)+\sum_n \delta \mathbb{U}^{(n)}(T)$ and $ \delta \mathbb{U}^{(n)}(T)$ is the term of order $W^n$ in the perturbative expansion. 

\begin{figure}
\begin{center}
\includegraphics[width=90mm,angle=0,clip]{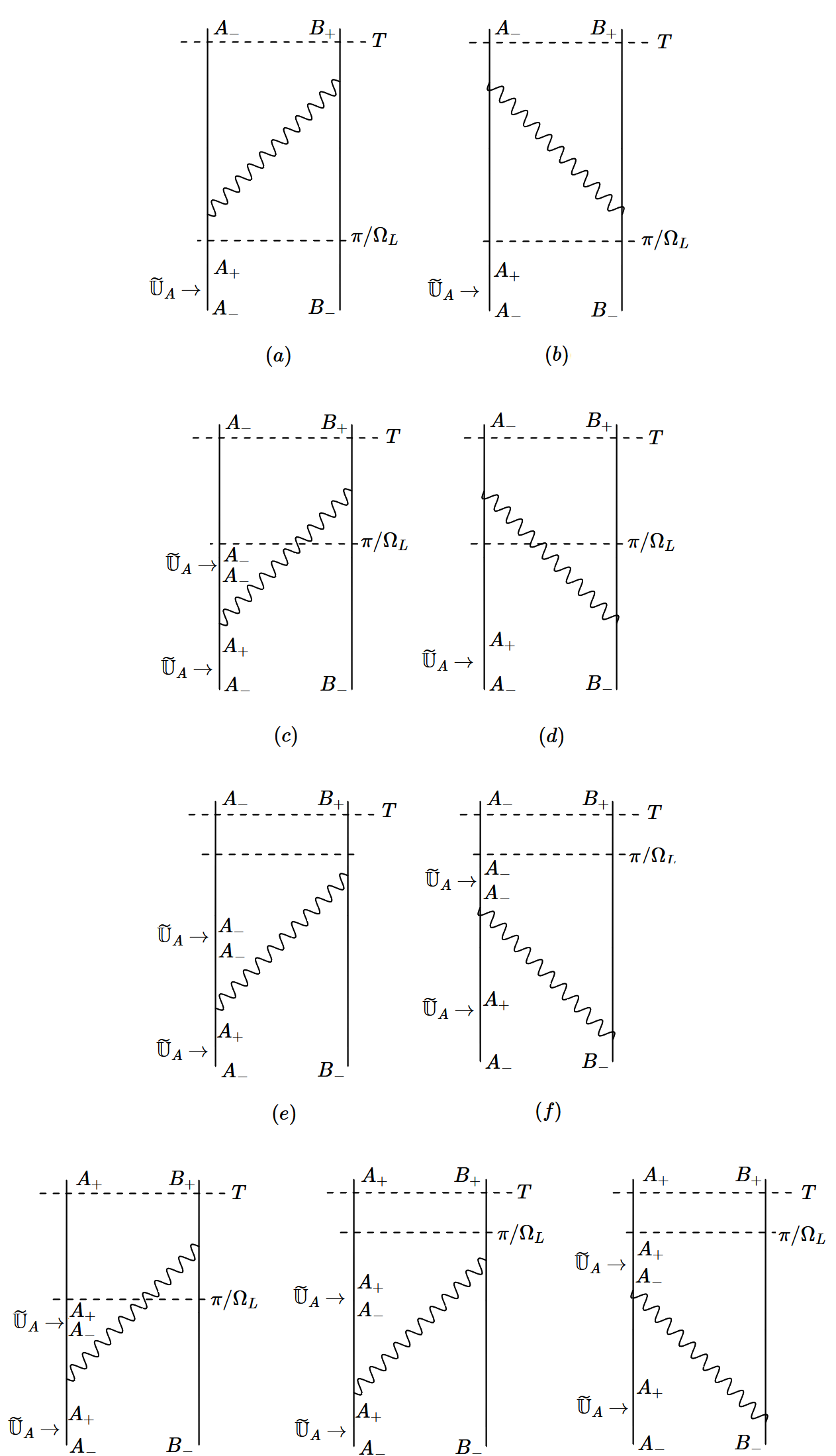}
\caption{Diagrammatic representation of the terms contributing to $\delta \mathbb{U}^{(2)}(T)$ in Eq.(\ref{laeq}). The action of the external laser is represented by horizontal arrows acting upon the atom $A$.}\label{fig6}
\end{center}
\end{figure}

In what follows, we evaluate the terms contributing to $\mathbb{U}(T)$. The leading ones, we will see, are those which are fully resonant and involve intermediate states with zero or one virtual photon. As advanced previously, they give rise to recurrent and summable series, being sufficient to compute the terms of orders $W^{2}$ and $W^{4}$, for an odd and an even number of exchanged photons, respectively. The former are depicted by the diagrams of Fig.\ref{fig6}. Their contributions are summarized in the expression, 
\begin{align}\label{laeq}
	&\delta \mathbb{U}^{(2)}(T)=-\hbar^{-2} \int_{0}^{\infty}\frac{\mathcal{V}k^{2}\textrm{d}k}{(2\pi)^{3}}\int_{0}^{\infty}\frac{\mathcal{V}k^{'2}\textrm{d}k'}{(2\pi)^{3}}\int_{0}^{4\pi}\textrm{d}\Theta\int_{0}^{4\pi}\textrm{d}\Theta' \nonumber\\ &\times\left[\int_{\pi/{\Omega_{L}}}^{T}\textrm{d}t\int_{\pi/{\Omega_{L}}}^{t}\textrm{d}t'+\int_{\pi/{\Omega_{L}}}^{T}\textrm{d}t\int_{0}^{\pi/{\Omega_{L}}}\textrm{d}t'+\int_{0}^{\pi/{\Omega_{L}}}\textrm{d}t\int_{0}^{t}\textrm{d}t' \right]\nonumber\\ &\times\Bigl[\widetilde{\mathbb{U}}_{0}(T-t)\mathbf{d}_{B}\cdot\mathbf{E}_{\mathbf{k}'}^{(-)}(\mathbf{R}_{B}) \widetilde{\mathbb{U}}_0(t-t')\mathbf{d}_{A}\cdot\mathbf{E}_{\mathbf{k}'}^{(+)}(\mathbf{R}_{A})\widetilde{\mathbb{U}}_0(t')\nonumber\\
	&+\widetilde{\mathbb{U}}_{0}(T-t)\mathbf{d}_{A}\cdot\mathbf{E}_{\mathbf{k}'}^{(-)}(\mathbf{R}_{A}) \widetilde{\mathbb{U}}_0(t-t')\mathbf{d}_{B}\cdot\mathbf{E}_{\mathbf{k}'}^{(+)}(\mathbf{R}_{B})\widetilde{\mathbb{U}}_0(t')\Bigr].
\end{align} 

As for the element of $\mathbb{U}$ proportional to $|A_-,B_+\rangle \langle A_-,B_-|$, the addition of all the contributions of second order yield
\begin{align}
	\delta \mathbb{U}^{(2)}_{B_{+}}&(T)=  -e^{-i\omega_0 T-\Gamma_{0}T/2} \Bigl[ \left(T-\frac{\pi}{\Omega_L}+i\partial_{\omega}\right) \widetilde{\Omega}(kR)|_{\omega=\omega_0}\nonumber\\ &+\mathcal{O}(\widetilde{\Omega}(k_0R)/\Omega_L)+\mathcal{O}(\widetilde{\Omega}(\Omega_{L}R/c)/\omega_0)\nonumber\\ &+ \frac{i}{2} \int_{0}^{\infty} \frac{du}{\pi} \frac{u^2-\omega_0^2}{(u^2+\omega_0^2)^2}\widetilde{\Omega}(iuR) \Bigr]|A_-,B_+\rangle \langle A_-,B_-|.
\end{align}
where only the resonant terms proportional to $T$ are summable, and those of the orders $\mathcal{O}(\widetilde{\Omega}(k_0R)/\Omega_L)$ and $\mathcal{O}(\widetilde{\Omega}(\Omega_{L}R/c)/\omega_0)$ are negligibly small in the strong coupling regime. As for the off-resonant term, in the form of a frequency integral over imaginary frequencies, it amounts to an overall phase shift term in the two-atom wavefunction which is only relevant in the near field provided that $\omega_{0}^{2}R/\Gamma_{0}c\lesssim1$.

Concerning the $\mathbb{U}$ element proportional to $|A_+,B_-\rangle \langle A_-,B_-|$, the dominant contribution comes from resonant processes involving an even number of single photon intermediate states (see Fig.\ref{fig4}). A computation analogous to that for $\delta \mathbb{U}^{(2)}_{B_{+}}$, but at order $W^{4}$ in this case, yields
\begin{align}
	\delta \mathbb{U}^{(4)}_{A_{+}}(T)=&\frac{i}{2} e^{-i\omega_0 T-\Gamma_{0}T/2} (T-\frac{\pi}{\Omega_L}+i\partial_{\omega})^2 \widetilde{\Omega}^2(kR)|_{\omega=\omega_0}\nonumber\\
	&\times |A_+,B_-\rangle \langle A_-,B_-|.
\end{align}
Also contributing to the element $|A_+,B_-\rangle \langle A_-,B_-|$, at order $W^{4}$, off-resonant terms come from processes with two-photon intermediate states (see Fig.\ref{fig5}). As with the off-resonant terms of order $W^{2}$, they amount to an overall phase shift term in the two-atom wavefunction which is generally negligible except in the near field regime, in which case the strong coupling condition for the laser field would be violated.
	
Finally, the $\mathbb{U}$ element proportional $|A_+,B_+\rangle \langle A_-,B_-|$, at leading order, is provided by the processes depicted by the diagrams of Figs.\ref{fig6}(g), \ref{fig6}(h) and \ref{fig6}(i), which yield terms of the order of $\widetilde{\Omega}(k_{0}R)/\Omega_{L}$ and thus, negligible in the strong coupling regime.  

The fully resonant terms involving the alternate exchange of either odd numbers or even numbers of single photons between the atoms in states $|A_+,B_-\rangle$  $|A_-,B_+\rangle$ can be summed up, amounting to $\mathbb{U}_{B_{+}}(T)$ and $ \mathbb{U}_{A_{+}}(T)$ in either case. That is, up to terms order $\tilde{\Omega}(k_{0}R)/\Omega_{L}$,
\begin{align}
	\mathbb{U}_{A+}(T)&=-i e^{-i\omega_0 T-\Gamma_{0}T/2} \cos\left[(T+i\partial_{\omega})\widetilde{\Omega}(kR)\right]_{\omega=\omega_0}\nonumber\\ 
	&\times |A_+,B_-\rangle \langle A_-,B_-|\label{Ua}\\
	\mathbb{U}_{B_+}(T)&=-e^{-i\omega_0 T-\Gamma_{0}T/2} \sin\left[(T+i\partial_{\omega})\widetilde{\Omega}(kR)\right]_{\omega=\omega_0}\nonumber\\ &\times|A_-,B_+\rangle\langle A_-,B_-|.\label{Ub} 
\end{align}
In these expressions we neglect the aforementioned phase-shift factors from off-resonant processes, and we note that they coincides with the result of the sudden approximation up to a phase shift $\pi$. As for the probability of finding the atomic system excited we get, by simply squaring the above elements,
\begin{align}
	P_{A_+}(T)&=\frac{e^{-\Gamma_0 T}}{2} [\cosh\left(2\Gamma_{kR}T-2\partial_{\omega}\Omega_{kR}\right)\nonumber\\&+\cos\left(2\Omega_{kR}T+2\partial_{\omega}\Gamma_{kR}\right)]_{\omega=\omega_0}\label{PA}\\	
	P_{B_+}(T)&=\frac{e^{-\Gamma_0 T}}{2} [\cosh\left(2\Gamma_{kR}T-2\partial_{\omega}\Omega_{kR}\right)\nonumber\\&-\cos\left(2\Omega_{kR}T+2\partial_{\omega}\Gamma_{kR}\right)]_{\omega=\omega_0}\label{PB}
\end{align}

\subsection{Photon states. Spontaneous emission}

Once equipped with the $\mathbb{U}$ elements which govern the atomic dynamics, we compute the element of $\mathbb{U}$ for the emission of single photons of any momentum and polarization , $\mathbb{U}_{\gamma}$, from the excited 
states of the atomic system. That is, the element for the transition from the initial state $|A_-,B_-; 0_\gamma\rangle$ to the states $|A_-,B_-;\gamma_{\mathbf{k},\boldsymbol{\epsilon}}\rangle$, with $\mathbf{k}$ and $\boldsymbol{\epsilon}$ being any momentum and polarization, passing through the intermediate states  $|A_-,B_+\rangle$ or $|A_+,B_-\rangle$. It reads, in terms of the $\mathbb{U}$ elements $\mathbb{U}_{B_+}$, $\mathbb{U}_{A_+}$, for a photon of momentum $\mathbf{k}$ and polarization $\boldsymbol{\epsilon}$
\begin{align}
	\mathbb{U}_{\gamma_{\mathbf{k},\boldsymbol{\epsilon}}}(T)&=-i \hbar^{-1} \int_0^T dt\mathbb{U}_0(T-t) W_A(t) \mathbb{U}_{A_+}(t)\nonumber\\  
	&-i \hbar^{-1} \int_0^T dt\mathbb{U}_0(T-t) W_B(t) \mathbb{U}_{B+}(t)\nonumber\\ 
	&= -i \hbar^{-1} \int_0^T dt|A_-,B_-,\gamma_{\mathbf{k},\boldsymbol{\epsilon}}\rangle e^{-i \omega (T-t)}\nonumber\\
	&\times\langle A_-,B_-,\gamma_{\mathbf{k},\boldsymbol{\epsilon}}|W_A(t)\mathbb{U}_{A_+}(t)\nonumber\\
	&-i \hbar^{-1} \int_0^T dt|A_-,B_-,\gamma_{\mathbf{k},\boldsymbol{\epsilon}}\rangle e^{-i \omega (T-t)}\nonumber\\
	&\times\langle A_-,B_-,\gamma_{\mathbf{k},\boldsymbol{\epsilon}}|W_B(t)\mathbb{U}_{B_+}(t).\label{Ug}
\end{align}

\begin{figure}
	\begin{center}
		\includegraphics[width=80mm,angle=0,clip]{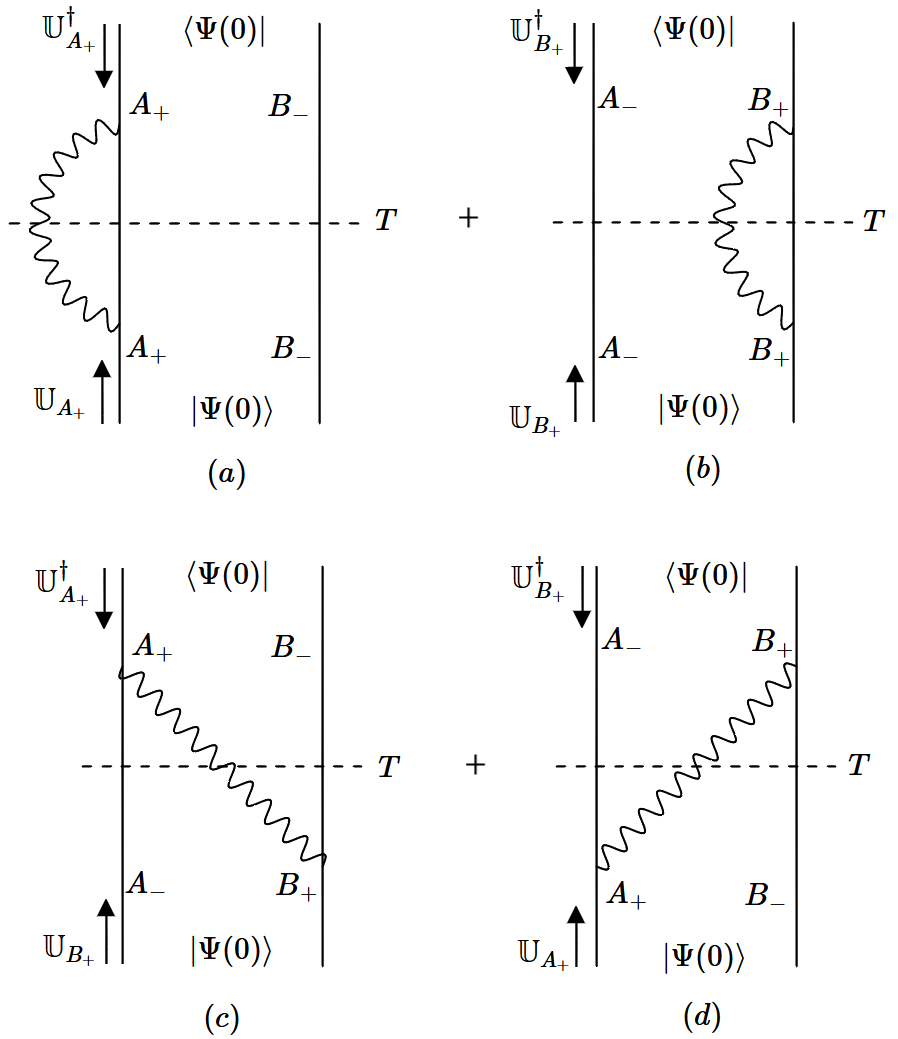}
		\caption{Diagrammatic representation of the spontaneous emission emission processes.}\label{figi}
	\end{center}
\end{figure}

From the above expression,  the probability of spontaneous emission of photons of any momentum and polariztion is 
\begin{align}
&P_{\gamma}(T)=\int\frac{d^{3}k}{(2\pi)^{3}}\mathcal{V}\sum_{\boldsymbol{\epsilon}}|\langle A_-,B_-;\gamma_{\mathbf{k},\boldsymbol{\epsilon}}|\mathbb{U}_{\gamma_{\mathbf{k},\boldsymbol{\epsilon}}}(T)|A_-,B_-;0_\gamma\rangle|^{2}\nonumber\\
&=\cosh\left(2 \partial_{\omega}\Omega_{k R}\right)-e^{-\Gamma_0 T}\cosh\left(2\Gamma_{kR} T- 2\partial_{\omega}\Omega_{k R}\right)\Bigr|_{\omega=\omega_0}.\label{Gtotal}
\end{align}
This expression implies processes in which the photons to be emitted are created and annihilated at the same atom, either $A$ or $B$ --those depicted by the diagrams of Figs.\ref{figi}(a) and \ref{figi}(b), and processes in which the photons are created and annihilated in different atoms --diagrams of Figs.\ref{figi}(c) and \ref{figi}(d). 

It is the latter the ones that give rise to the directionality of the spontaneous emission that we will address in Sec.\ref{directionality}. In the near field limit, $k_0 R \ll 1$, we find 
\begin{equation}
P_{\gamma}(T)\simeq\frac{(1-e^{-2\Gamma_0 T})}{2}e^{2 \partial_{\omega}\Omega_{k R}|_{\omega=\omega_0}} \sim \frac{1-e^{-2\Gamma_0 T}}{2}.\label{Gnear}
\end{equation}

In view of the above results two comments are in order. In the first place, from Eq.(\ref{Gnear}) we see that, in contrast to the subradiant Dicke state, in the near field, the spontaneous emission induced by the electric dipole coupling is only half-inhibited. Second, adding up the probability of emission in Eq.(\ref{Gtotal}) and those for the atomic excitation of Eqs.(\ref{PA}) and (\ref{PB}), we get
\begin{equation}
		P_{A_+}(T)+	P_{B_+}(T)+	P_{\gamma}(T)=\cosh\left(2 \partial_{\omega}\Omega_{k R}\right)_{\omega=\omega_0}.
\end{equation}
Unitarity demands this probability to be unity. It can be verified that either in the near field or in the far field regimes the above probability is indeed unity up to terms of the order of $(k_{0}\mu^{2}/e^{2}R)^2$ or $k_{0}^{4}\mu^{4}/e^{4}$ in either case, both much less than unity in many orders of magnitude.

\section{Atomic forces and directional emission}\label{IV}

In this section we compute the forces upon each atom paying special attention to their non-reciprocal and non-conservative components. To these ends, we extend the perturbative treatment developed in Refs.\cite{Julio_PRA1,Julio_PRA2} for the case of dissimilar atoms and weak interacting alike atoms.

Our atomic system possesses a conserved total momentum  \cite{Cohen_Kawkathesis_KawkavanTiggelen_Dippel,Kawkathesis}, $[H,\mathbf{K}]=\mathbf{0}$, 
\begin{equation}
	\mathbf{K}=\mathbf{P}_{A}+\mathbf{P}_{B}+\mathbf{P}_{\perp}^{\gamma},\label{K}
\end{equation}
where $\mathbf{P}_{A,B}$ are the  canonical conjugate momenta of the centers of mass of each atom and $\mathbf{P}_{\perp}^{\gamma}=\sum_{\mathbf{k},\boldsymbol{\epsilon}}\hbar\mathbf{k}\:a^{\dagger}_{\mathbf{k},\boldsymbol{\epsilon}}a_{\mathbf{k},\boldsymbol{\epsilon}}$ 
is the transverse EM momentum. Further, the canonical conjugate 
momenta can be written as 
\begin{equation}
	\mathbf{P}_{A}+\mathbf{P}_{B}=m_{A}\dot{\mathbf{R}}_{A}+m_{B}\dot{\mathbf{R}}_{B}-\mathbf{d}_{A}\times\mathbf{B}(\mathbf{R}_{A})-\mathbf{d}_{B}\times\mathbf{B}(\mathbf{R}_{B}),\label{K}
\end{equation}
where the first two terms are the kinetic momenta of the centers of mass of each atom, and the rest of the terms account for the longitudinal EM momentum in the electric dipole approximation \cite{,Cohen, Cohen_Kawkathesis_KawkavanTiggelen_Dippel}, i.e., the so-called R\"ontgen momentum \cite{Baxter1,Baxter2}, which are proportional to the magnetic fields at the positions of each atom, $\mathbf{B}(\mathbf{R}_{A,B})=\boldsymbol{\nabla}_{A,B}\times\mathbf{A}(\mathbf{R}_{A,B})$.

Following Refs.\cite{MeQFriction,My_Net_PRA,MePRAvdW,Julio_PRA1,Julio_PRA2}, the force on each atom is computed applying the time derivative to the expectation value of the kinetic momenta 
of the centers of mass of each atom. Writing the latter in terms of the canonical conjugate momenta and the longitudinal EM momentum, we arrive at
\begin{align}
	&\langle\mathbf{F}_{A,B}(T)\rangle=\partial_{T}\langle m_{A,B}\dot{\mathbf{R}}_{A,B}\rangle_{T}\label{Force}\\
	&=-i\hbar\partial_{T}\langle\Psi(0)|\mathbb{U}^{\dagger}(T)\boldsymbol{\nabla}_{A,B}\mathbb{U}(T)|\Psi(0)\rangle\nonumber\\
	&+\partial_{T}\langle\Psi(0)|\mathbb{U}^{\dagger}(T)\mathbf{d}_{A,B}\times\mathbf{B}(\mathbf{R}_{A,B})\mathbb{U}(T)|\Psi(0)\rangle\nonumber\\
	&=-\langle \boldsymbol{\nabla}_{A,B}W_{A,B}(T)\rangle+
	\partial_{T}\langle[\mathbf{d}_{A,B}\times\mathbf{B}(\mathbf{R}_{A,B})](T)\rangle,\nonumber
\end{align}
The first term on the \emph{r.h.s.} of last equality is a conservative force along the interatomic axis.
The second term is a non-conservative force equivalent to the time derivative of the longitudinal EM momentum at each atom, with opposite sign.

We will express the conservative forces in terms of the Green tensor of Eq.(\ref{GpaE}). Likewise, for the nonconservative forces we will use the relation of Eq.(\ref{GpaB}).

\subsection{Conservative forces}
Once computed the elements of the time evolution propagator for the atomic dynamics, the computation of the dominant forces on each atom reduces to the calculation of the expectation values of their operators at leading order in $W$, using $\mathbb{U}$ as the non-perturbative propagator.
\begin{figure}
	\begin{center}
		\includegraphics[width=90mm,angle=0,clip]{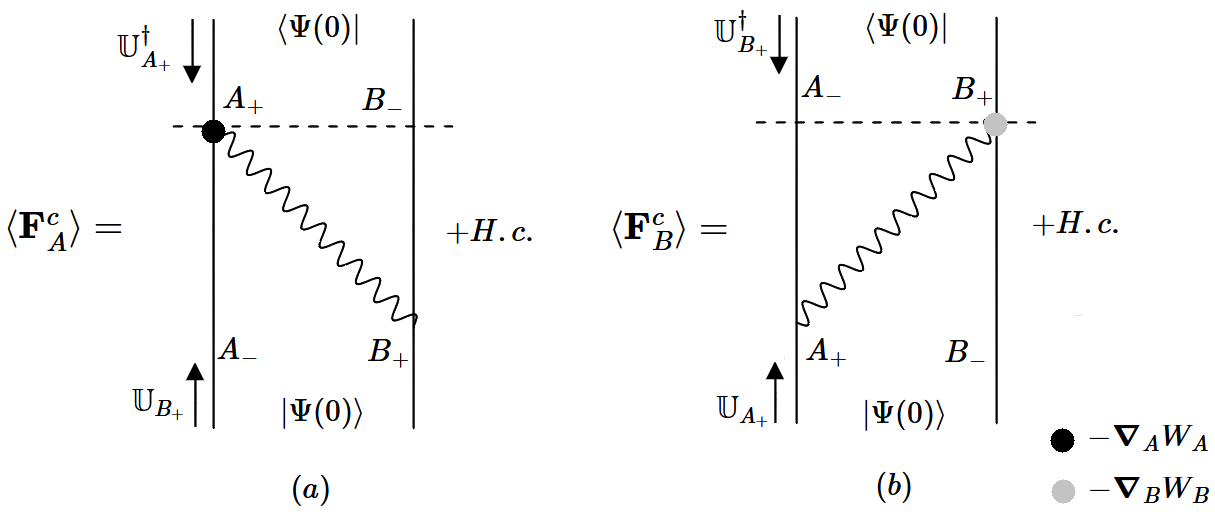}
		\caption{Diagrammatic representation of the processes contributing to the resonant conservative forces.}\label{figii}
	\end{center}
\end{figure}

\begin{figure}
	\begin{center}
		\includegraphics[width=90mm,angle=0,clip]{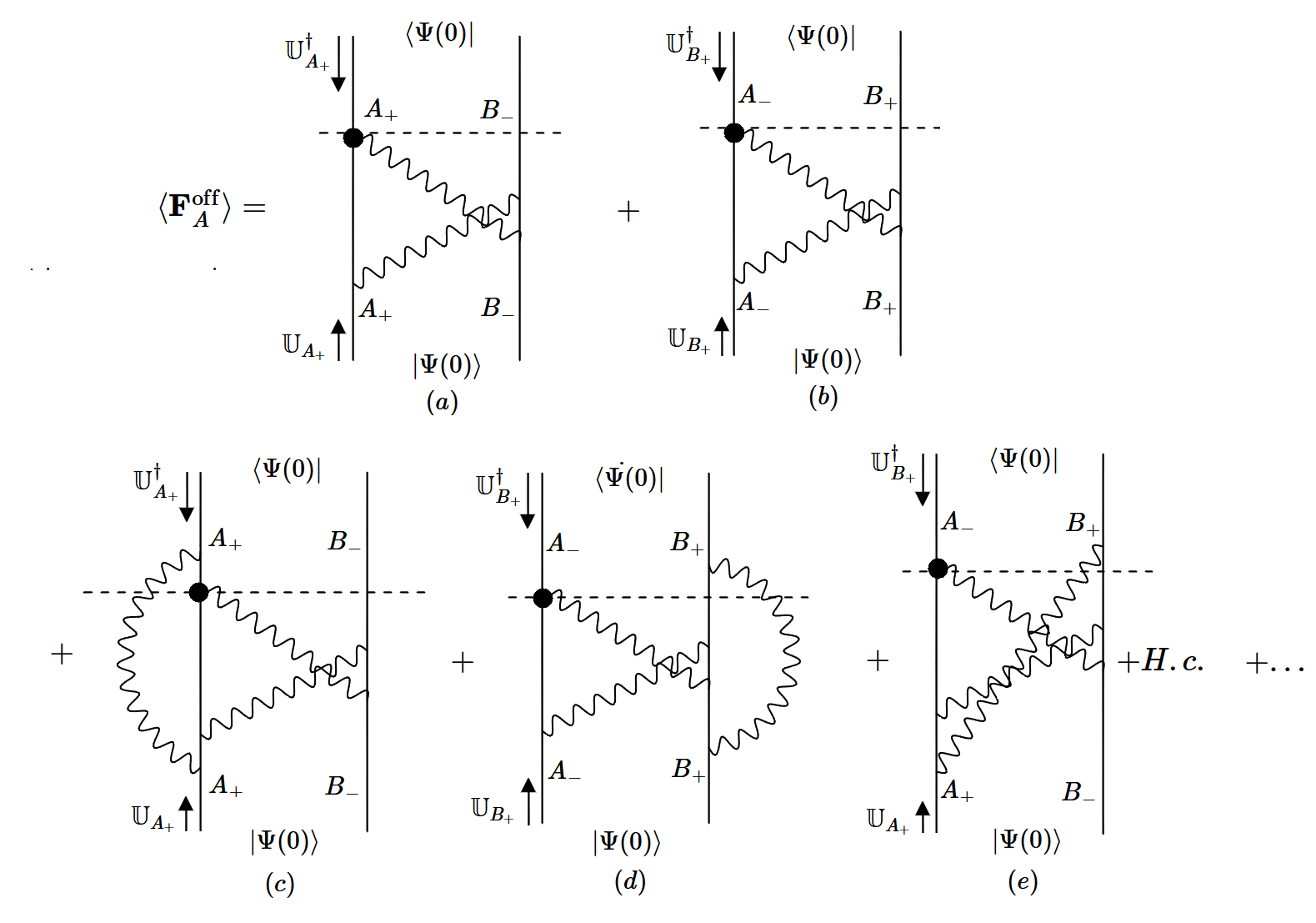}
		\caption{Diagrammatic representation of some of the processes contributing to the off-resonant vdW forces, with two-atom [diagrams (a) and (b)] and three-atom [diagrams (c), (d), (e)] intermediate states. In the former ones, any of the atoms is excited, while in the latter ones an actual photon has been emitted at the observation time $T$.}\label{figiii}
	\end{center}
\end{figure}

The corresponding diagrams are represented in Figs.\ref{figii}(a) and \ref{figii}(b) for the resonant forces, and Fig.\ref{figiii} for the off-resonant forces. As for the former, which are the dominant ones, they  read, 
\begin{align}
	\langle \mathbf{F}^{c}_{A}(T) \rangle&= -2 \text{Re}\big\{-i \hbar^{-1}\langle \Psi(0)| \mathbb{U}_{A_+}^{\dagger}(T)\boldsymbol{\nabla}_{A}W_{A}\nonumber\\ 
	&\times\int_{0}^{T} dt \mathbb{U}_0(T-t)W_{B}(t)\mathbb{U}_{B_+}(t) |\Psi(0) \rangle\big\}\nonumber\\
	&=\hbar e^{- \Gamma_0 T} \Bigl[\sin  \left[2(\Omega_{kR} T+ \partial_{\omega} \Gamma_{kR}) \right]\boldsymbol{\nabla}_{\mathbf{R}}\Gamma_{kR}\nonumber\\
	&+ \sinh \left[2(\Gamma_{kR} T-\partial_{\omega}\Omega_{kR}) \right]\boldsymbol{\nabla}_{\mathbf{R}} \Omega_{kR} \Bigr]_{\omega=\omega_0},\label{FA}
\end{align}
\begin{align}
	\langle \mathbf{F}^{c}_{B}(T) \rangle&= -2 \text{Re}\big\{-i \hbar^{-1}\langle \Psi(0)| \mathbb{U}_{B+}^{\dagger}(T)\boldsymbol{\nabla}_{B}W_{B}\nonumber\\ 
	&\times \int_{0}^{T} dt\mathbb{U}_0(T-t)W_{A}(t)\mathbb{U}_{A_+}(t) |\Psi(0) \rangle\big\}\nonumber\\
	&=\hbar e^{- \Gamma_0 T} \Bigl[\sin  \left[2(\Omega_{kR} T+ \partial_{\omega} \Gamma_{kR}) \right]\boldsymbol{\nabla}_{\mathbf{R}}\Gamma_{kR}\nonumber\\
	&- \sinh \left[2(\Gamma_{kR} T-\partial_{\omega}\Omega_{kR}) \right]\boldsymbol{\nabla}_{\mathbf{R}} \Omega_{kR} \Bigr]_{\omega=\omega_0}.\label{FB}
\end{align}

The asymmetry and non-stationarity of the atomic states breaks dynamically the parity symmetry of the atomic system, which results in  non-reciprocal forces. That implies the existence of a net oscillatory internal force upon the whole system,  
\begin{align}
		\langle\mathbf{F}^{c}_A(T)+\mathbf{F}^{c}_B(T)\rangle&=2\hbar  e^{- \Gamma_0 T}\sin  \left[2(\Omega_{kR} T+ \partial_{\omega} \Gamma_{kR}) \right]\nonumber\\
		&\times\boldsymbol{\nabla}_{\mathbf{R}}\Gamma_{kR}|_{\omega=\omega_0}.\label{Ftotal}
\end{align}

For the sake of completeness, for the computation of the off-resonant conservative forces, it suffices to consider the processes up to order $W^{6}$. They involve terms of order $W^{4}$ in which the state of the system is either $|A_+,B_-;0_{\gamma}\rangle$-- i.e., that in Fig.\ref{figiii}(a), which corresponds to the aforementioned phase-shift of the two-atom wave function provided by the $\mathbb{U}$ element $|A_+,B_-\rangle \langle A_-,B_-|$, or $|A_-,B_+;0_{\gamma}\rangle$-- i.e., that in Fig.\ref{figiii}(b) . The terms of order $W^{6}$ are those in which the state of the system is $|A_-,B_-;\gamma_{\mathbf{k},\boldsymbol{\epsilon}}\rangle$, i.e. , those in Figs.\ref{figiii}(c), \ref{figiii}(d) and \ref{figiii}(e), in which an actual photon has been emitted from the atomic system at the observation time $T$ \cite{Julio_PRA1}.
 The computation is analogous to that in Ref.\cite{Julio_PRA1} considering the identical atoms limit. Adding up the contribution of all of them we obtain 
\begin{align}
	\langle \mathbf{F}^{\text{off}}_{A,B}(T)\rangle& =\pm \frac{4\omega_{0}^{2}}{c^3\epsilon_{0}^{2}\hbar}\int_{0}^{\infty}\frac{dq}{\pi}
	\frac{q^4 \boldsymbol{\mu}_{A}\cdot\mathbb{G}(iqR)\cdot\boldsymbol{\mu}_{B} }{(q^2+k_0^2)^{2}}\nonumber\\
	&\times\boldsymbol{\nabla}_{\mathbf{R}}[\boldsymbol{\mu}_{B}\cdot\mathbb{G}(iqR)\cdot\boldsymbol{\mu}_{A}] \Bigl[\cosh\left(2 \partial_{\omega}\Omega_{k R}\right)_{\omega=\omega_0}\nonumber\\
	&-2e^{-\Gamma_0 T}\cosh\left(2\Gamma_{kR} T- 2\partial_{\omega}\Omega_{k R}\right)\Bigr]_{\omega=\omega_0}.\label{Foff}
\end{align}
In this expression it is straightforward to identify 
\begin{align}
		\langle \mathbf{F}^{\text{off}}_{A,B}(T)\rangle& =\left [P_{A_+}(T)+ P_{B_+}(T)\right]\nonumber\\
		&\times\langle A_{+,-},B_{-,+}|-\boldsymbol{\nabla}_{A,B}W_{A,B}| A_{+,-},B_{-,+}\rangle\nonumber\\
		&+P_{\gamma}(T)\langle A_-,B_-;\gamma_{k_{0}}|-\boldsymbol{\nabla}_{A,B}W_{A,B}|A_-,B_-;\gamma_{k_{0}}\rangle\nonumber
\end{align}
In contrast to the resonant ones, off-resonant forces are reciprocal.

\subsection{Non-conservative forces}
\begin{figure}
	\begin{center}
		\includegraphics[width=90mm,angle=0,clip]{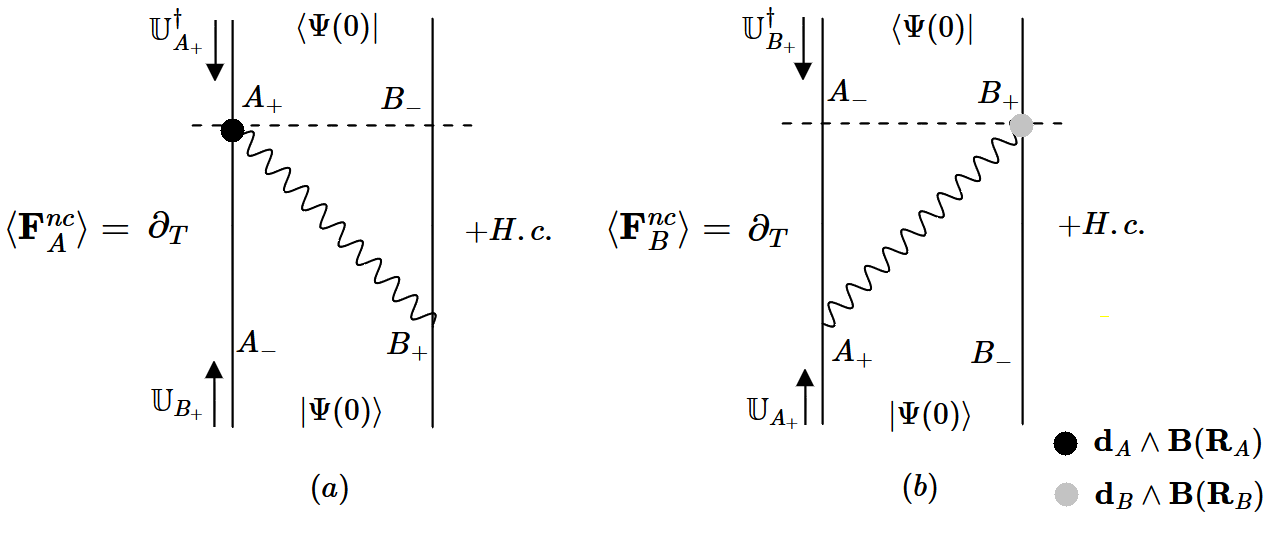}
		\caption{Diagrammatic representation of the processes contributing to the nonconservative forces.}\label{figiv}
	\end{center}
\end{figure}

These forces correspond to the time variation of the longitudinal EM momentum. Again, their dominant contribution comes from the resonant terms depicted in Fig.\ref{figiv}, i.e.,
 
\begin{align}
	\langle \mathbf{F}^{nc}_{A}(T) \rangle&= 2\partial_{T}\text{Re}\big\{-i \hbar^{-1}\langle \Psi(0)| \mathbb{U}_{A_+}^{\dagger}(T)\mathbf{d}_{A}\times\mathbf{B}(\mathbf{R}_{A})\nonumber\\ 
	&\times\int_{0}^{T} dt \mathbb{U}_0(T-t)W_{B}(t)\mathbb{U}_{B_+}(t) |\Psi(0) \rangle\big\},
\end{align}
\begin{align}
	\langle \mathbf{F}^{nc}_{B}(T) \rangle&= 2\partial_{T}\text{Re}\big\{-i \hbar^{-1}\langle \Psi(0)| \mathbb{U}_{B+}^{\dagger}(T))\mathbf{d}_{B}\times\mathbf{B}(\mathbf{R}_{B})\nonumber\\ 
	&\times \int_{0}^{T} dt\mathbb{U}_0(T-t)W_{A}(t)\mathbb{U}_{A_+}(t) |\Psi(0) \rangle\big\}.
\end{align}
We write their full expressions below, in terms of the tensors 
\begin{align}
\boldsymbol{\Lambda}_{AB}&=-\boldsymbol{\mu}_A \times \boldsymbol{\nabla} \times  \text{Im}\mathbb{G}(kR) \cdot \boldsymbol{\mu}_B,\nonumber\\
\boldsymbol{\Sigma}_{AB}&=\boldsymbol{\mu}_A \wedge \boldsymbol{\nabla} \wedge  \text{Re}\mathbb{G}(kR) \cdot \boldsymbol{\mu}_B,
\end{align}

	\begin{equation}
	\begin{split}
		\langle \mathbf{F}_A^{nc}(T)\rangle&=\frac{2 e^{-\Gamma_0T}}{\epsilon_0 k_0 c}\Bigl[\cos\left[2(\Omega_{kR} T+\partial_{\omega} \Gamma_{kR})\right]\boldsymbol{\Lambda}_{AB} \Omega_{kR}\\
		&- \cosh\left[2(\Gamma_{kR} T-\partial_{\omega}\Omega_{kR})\right]\boldsymbol{\Sigma}_{AB} \Gamma_{kR}\Bigr]_{\omega=\omega_0}\\
		&-\frac{\Gamma_0 e^{-\Gamma_0T}}{\epsilon_0 k_0 c}\Bigl[
		\sin\left[2(\Omega_{kR} T+\partial_{\omega} \Gamma_{kR})\right]\boldsymbol{\Lambda}_{AB}\\
		&-\sinh\left[2(\Gamma_{kR} T-\partial_{\omega}\Omega_{kR})\right]\boldsymbol{\Sigma}_{AB}\Bigr]_{\omega=\omega_0},\label{FncA}
	\end{split}
\end{equation}	
\begin{equation}
	\begin{split}
		\langle \mathbf{F}_B^{nc}(T)\rangle&=-\frac{2 e^{-\Gamma_0T}}{\epsilon_0 k_0 c}\Bigl[\cos\left[2(\Omega_{kR} T+\partial_{\omega} \Gamma_{kR})\right]\boldsymbol{\Lambda}_{AB} \Omega_{kR}\\
		&+ \cosh\left[2(\Gamma_{kR} T-\partial_{\omega}\Omega_{kR})\right]\boldsymbol{\Sigma}_{AB} \Gamma_{kR}\Bigr]_{\omega=\omega_0}\\
		&+\frac{\Gamma_0 e^{-\Gamma_0T}}{\epsilon_0 k_0 c}\Bigl[
		\sin\left[2(\Omega_{kR} T+\partial_{\omega} \Gamma_{kR})\right]\boldsymbol{\Lambda}_{AB}\\
		&+\sinh\left[2(\Gamma_{kR} T-\partial_{\omega}\Omega_{kR})\right]\boldsymbol{\Sigma}_{AB}\Bigr]_{\omega=\omega_0}.\label{FncB}
	\end{split}
\end{equation}
Once more, these forces contain non-reciprocal components which amount to a net non-coservative force upon the whole system,
\begin{align}
		&\langle \mathbf{F}_A^{nc}(T)+ \mathbf{F}_B^{nc}(T)\rangle=\frac{-4 e^{-\Gamma_0T}}{\epsilon_0 k_0 c}\Bigl\{\cosh\left[2(\Gamma_{kR} T-\partial_{\omega}\Omega_{kR})\right]\nonumber\\
		&\times\boldsymbol{\Sigma}_{AB} \Gamma_{kR}
		-\frac{\Gamma_0}{2}\sinh\left[2(\Gamma_{kR} T-\partial_{\omega}\Omega_{kR})\right]\boldsymbol{\Sigma}_{AB}\Bigr\}_{\omega=\omega_0}.\label{Fnctotal}
\end{align}	

\subsection{Directional emission}\label{directionality}

In the previous Section we computed the total probability of single photon emission. Here we are interested in the directional nature of that emission. Hence, we will show, it is related to the non-reciprocal character of the conservative forces outlined above. 

From the equations (\ref{Ug}) and (\ref{Gtotal}) for the total emission probability, integrating the latter in frequencies and deriving with respect to time, we compute the rate of photon emission per unit of solid angle,
\begin{align}
\frac{d \Gamma}{d \Theta_{\mathbf{k}_{0}}}&=\partial_{T}\int\frac{\mathcal{V}k^{2}dk}{(2\pi)^{3}}\sum_{\boldsymbol{\epsilon}}|\langle A_-,B_-;\gamma_{\mathbf{k},\boldsymbol{\epsilon}}|\mathbb{U}_{\gamma_{\mathbf{k},\boldsymbol{\epsilon}}}(T)|A_-,B_-;0_\gamma\rangle|^{2}\nonumber\\
&=- \frac{k_0^3 \boldsymbol{\mu}_A\cdot(\mathbb{I}-\hat{\mathbf{k}} \otimes \hat{\mathbf{k}})\cdot\boldsymbol{\mu}_B e^{-\Gamma_0 T} }{2(2 \pi \epsilon_0)^2}\nonumber\\
&\times  \Biggl\{\cos (kR \cos \theta) \sinh(2 \Gamma_{kR}T-2 \partial_{\omega}\Omega_{k R})\nonumber\\
&+\sin (kR \cos \theta)\sin(2 \Omega_{kR}T-2 \partial_{\omega}\Gamma_{k R})\Biggr\}_{\omega=\omega_{0}}.
\end{align}
From the above expression we see that the term proportional to $\sin (k_0R \cos \theta)$ is the one which causes the asymmetry of the emission. Correspondingly, there exists a net time-variation of the momentum of the emitted photons,
\begin{align}
	\langle\dot{\mathbf{P}}_{\gamma}&(T)\rangle=\partial_{T}\int\frac{\mathcal{V}d^{3}k}{(2\pi)^{3}}\mathbf{k}\nonumber\\
	&\times\sum_{\boldsymbol{\epsilon}}|\langle A_-,B_-;\gamma_{\mathbf{k},\boldsymbol{\epsilon}}|\mathbb{U}_{\gamma_{\mathbf{k},\boldsymbol{\epsilon}}}(T)|A_-,B_-;0_\gamma\rangle|^{2}\nonumber\\
	&=-2\hbar  e^{- \Gamma_0 T}\sin  \left[2(\Omega_{kR} T+ \partial_{\omega} \Gamma_{kR}) \right]\boldsymbol{\nabla}_{\mathbf{R}}\Gamma_{kR}|_{\omega=\omega_0}.\label{Pconserved}
\end{align}
which is equivalent to the net conservative force upon the system, with opposite sign, $\langle\dot{\mathbf{P}}_{\gamma}(T)\rangle=-\langle\mathbf{F}^{c}_A(T)+\mathbf{F}^{c}_B(T)\rangle$, in accord with total momentum conservation.

\section{Conclusions and discussion}\label{V}
 Within the framework of the wavefunction formalism and applying diagrammatic perturbative techniques to time-dependent Hamiltonian-QED, closed-form expressions have been derived for several functions and observables that describe the dynamics of a binary system of identical two-level atoms which share an excitation. These are, the  wavefunction of the system [Eqs.(\ref{Ua}), (\ref{Ub}) and (\ref{Ug}) ], the atomic population levels [Eqs.(\ref{PA}) and (\ref{PB})], and the probability of spontaneous emission [Eq.(\ref{Gtotal})]. From Eq.(\ref{Gnear}) we see that, in contrast to the subradiant Dicke state, in the near field, the spontaneous emission induced by the electric dipole coupling is only half-inhibited. In regards to the dynamics of the atomic system, this is governed by resonant forces which contain both conservative [Eqs.(\ref{FA})) and (\ref{FB})] and non-conservative components [Eqs.(\ref{FncA}) and (\ref{FncB})]. As for the former, they are made of two kinds of terms, namely, those that oscillate in time at the frequency of the interatomic interaction, $2\Omega_{k_{0}R}$, and whose strength is proportional to the gradient of the collective spontaneous emission rate, $\boldsymbol{\nabla}_{\mathbf{R}}\Gamma_{k_{0}R}$; and those which are attenuated by $\Gamma_{k_{0}R}$ and are proportional to $\boldsymbol{\nabla}_{\mathbf{R}}\Omega_{k_{0}R}$ instead. The former ones are non-reciprocal, which results in a net internal force upon the atomic system [Eq.(\ref{Ftotal})] and in an apparent violation of the Newtonian principle of action-reaction. The latter is interpreted in Sec.\ref{directionality} in terms of the directionality of the collective spontaneous emission, which results itself from the violation of the parity symmetry in the atomic system. Hence we have verified that the kinetic momentum gained by the atomic system equals, at any time and with opposite sign, the momentum gained by the EM field, in agreement with the conservation of total linear momentum --see Eq.(\ref{Pconserved}). It is worth mentioning here that there also exists a net non-conservative force upon the system [Eq.(\ref{Fnctotal})]. This force equals, with opposite sign, the time-variation of the so-called EM longitudinal momentum, thus guaranteeing again the conservation of total linear momentum.

We conclude our Discussion with an estimate of the displacement caused by the nonreciprocal forces on a binary system of Rydberg atoms, with the aim of motivating the design of an  experiment for its detection.  In order to facilitate their  observation, we assume that the reciprocal forces are counterbalanced by the external fields of the experimental setup --eg., those of optical lattices or optical tweezers. This allows us to consider that the interatomic distance between the atoms remains constant at any time and we focus ourselves on the displacement of the center of mass of the system. By integrating in time the expression for the net force in Eq.(\ref{Ftotal}), we find that the displacement of the center of the mass is
\begin{widetext}

\begin{align}
	S_{CM}= \frac{\hbar \boldsymbol{\nabla}_{\mathbf{R}} \Gamma_{k_0R}}{ \left(1+ \mathcal{R}^2\right)^2 M \Gamma_0^2} \left[\left(1+\mathcal{R}^2\right)2 \Omega_{k_0R}T-2 \mathcal{R} \left(1-e^{-\Gamma_0T}\cos\left(2 \Omega_{k_0R}T\right)\right)+(1-\mathcal{R}^2) e^{-\Gamma_0T}\sin \left(2 \Omega_{k_0R}T\right)\right],
\end{align}

\end{widetext}
where $M$ is the mass of a single atom and $\mathcal{R}=2 \Omega_{k_0R}/\Gamma_0$. For the case of circular Rydberg atoms, the transition dipole moments and the corresponding transition frequencies between two adjacent circular states scale with the principal quantum number $n$ as $\mu\approx e a_0 n^{2}$, $ \hbar \omega_{0}\approx E_0 n^{-3}$, respectively, where $e$ is the electron charge, $a_0$ the Bohr radius and $E_0$ the energy of the fundamental state of the Hydrogen atom \cite{Hare}.
This implies that $\langle \mathbf{F}^{c}_A(T)+\mathbf{F}^{c}_B(T) \rangle \sim n^{-8}$ and 
$S_{CM}\sim n^{2}$, i.e., while the net force decreases rapidly with $n$ for circular Rydberg atoms, the center of mass displacements
increases quadratically with $n$.

\begin{figure}[H]
	\begin{center}
		\includegraphics[width=85mm,angle=0,clip]{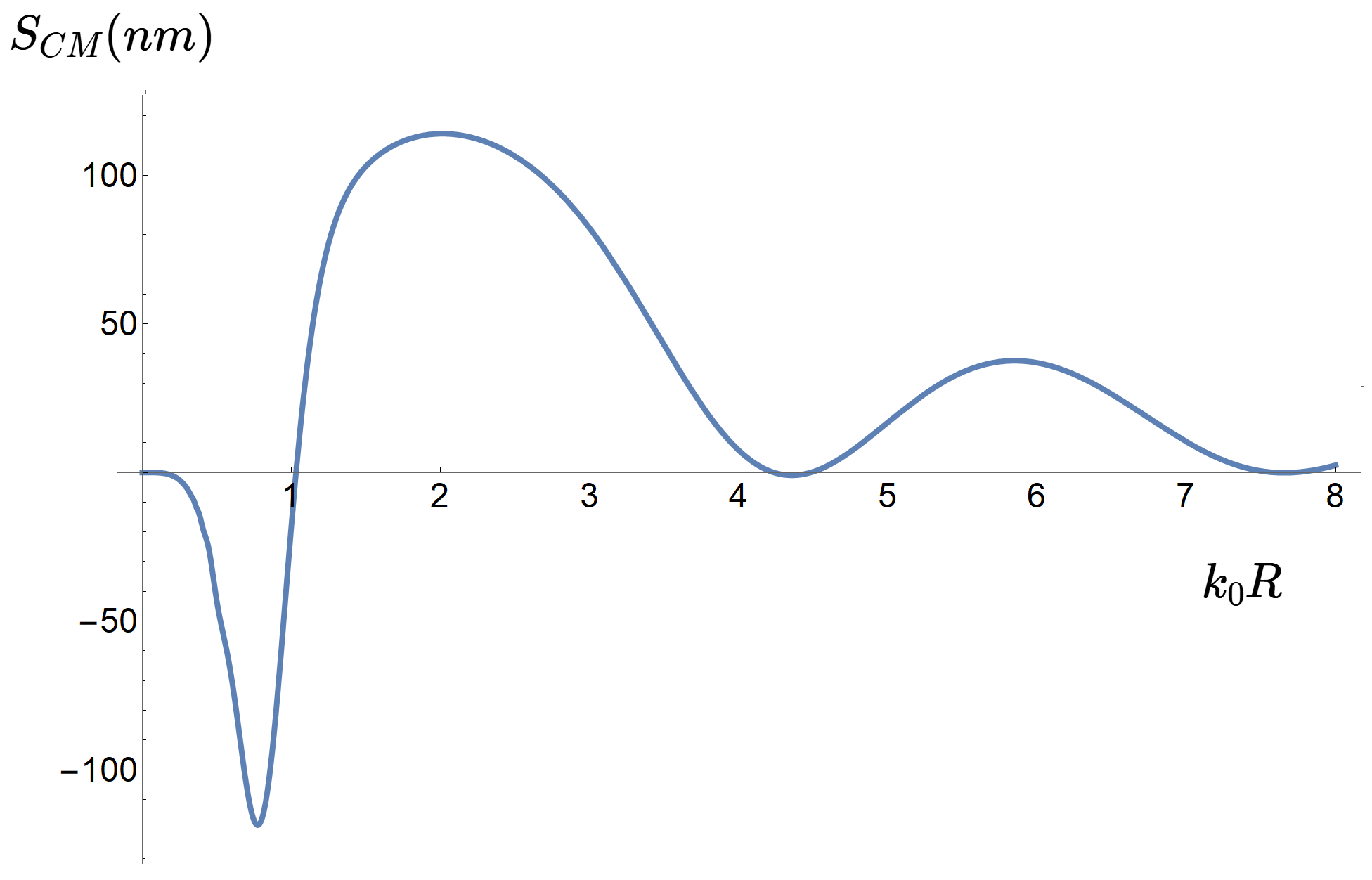}
		\caption{Graphical representation of the net displacement of the center of mass of a binary system of Li Rydberg atoms over a lifetime in terms of the interatomic distance. The initial state of the system is $|\Psi(0)\rangle=|A_+\rangle \otimes |B_-\rangle  =|70 C\rangle \otimes |69 C\rangle =|70,69,69\rangle \otimes |69,68,68\rangle $. Negative values of the displacement indicate movement towards atom $A$, while positive values indicate movement towards atom $B$.}\label{fig_displacement}
	\end{center}
\end{figure}

For the sake of illustration, let us consider the case of a binary system of lithium Rydberg atoms in initial circular states $|n C \rangle$ with $n \sim 70$, for which the coherent manipulation is feasible \cite{Zhang2010}. In particular, let us consider that one of the atoms is initially excited to the circular state $|A_+\rangle=|nC\rangle=|n,l,m_l \rangle= |70,69,69\rangle$ while the other is in the state $|B_-=|(n-1)C\rangle=|n-1,l,m_l \rangle= |69,68,68\rangle$. The quantization axis is defined by a director electrostatic field oriented perpendicular to the interatomic axis. Further, for the sake of simplicity, we consider the two-level atom approximation and restrict our calculation to the transition $|n C \rangle\leftrightarrow|(n-1)C\rangle$, with variations of the angular momentum $\Delta l= \pm 1$, $\Delta m_l=\pm 1$. The net displacement of the center of mass over a lifetime $T \approx \Gamma_{0}^{-1}$ is represented in Fig. \ref{fig_displacement} in terms of the normalized interatomic distance $k_0R$, with $\lambda_{0}=2\pi k_{0}=448\:\mu$m. We find displacements as large as   $S_{CM} \approx 120\:$nm for an interatomic distance $k_0R \approx 0.77$ and $k_0R \approx 2$. 
\vspace{-1.8cm}

\acknowledgments
J. S.-C. acknowledges the financial support from the project QCAYLE (NextGenerationEU funds, PRTRC17.I1) and from the Doctorate Program Funds No. UVa 2021 and Banco Santander.


\begin{thebibliography}{105}
\bibitem{comment} In this respect, forces related to the time variation of the longitudinal EM momentum have been 
also estimated in magnetochiral systems \cite{KawkavanTiggelenRikken,myPRLchiral,JPCM}.
\end{thebibliography}
\end{document}